# Simultaneous optical power delivery and distributed sensing through cross-band wavelength multiplexing over fiber link

Tianye Huang,[1,2,3,4,*] Lu Guo,[1] Xinyu Wang,[1] Yao Chen,[1] Jing Zhang,[1] Ming Zhu,[2] Mingkong Lu,[5] Kaifu Chen,[1] Hanlin Guo,[1] Liangming Xiong,[6] Xiangyun Hu,[1] and Perry Ping Shum,[7]

[1] School of Mechanical Engineering and Electronic Information, China University of Geosciences (Wuhan), Wuhan, Hubei, China
[2] Wuhan National Laboratory for Optoelectronics, Wuhan, Hubei, China
[3] Shenzhen Research Institute of China University of Geosciences, Shenzhen, Guangdong, China
[4] SINOPEC Key Laboratory of Geophysics, Nanjing, Jiangsu, China
[5] Yazheng Technology Group Co., Ltd, Wuhan, Hubei, China
[6] Yangtze optical fiber and cable Joint Stock Limited Company, Wuhan, Hubei, China
[7] Department of Electronic and Electrical Engineering, South University of Science and Technology, Shenzhen, Guangdong, China.

*Corresponding author: Tianye Huang (E-mail:*huangty@cug.edu.cn)

## Abstract

Optical fibers offer significant advantages in both power delivery and distributed sensing. In remote areas where stable power supply is not easy to access, the distributed optical fiber sensing (DOFS) which offers long distance monitoring capability and the power-over-fiber (PoF) which can provide energy for connected electronics or other sensors are highly desired simultaneously. In this letter, the PoF-DOFS hybrid system is proposed and experimentally verified for the first time. By multiplexing the power channel and sensing channel with large wavelength separation, the cross-talk is greatly reduced. The results show that the Brillouin frequency shift under different temperature in the Brillouin optical time domain reflectometry remains unaffected by the high-power transmission background and the power delivery efficiency up to ~66% can be achieved over 1.3 km fiber link. This work paves the way for further research on PoF-DOFS hybrid system and gives a valuable solution for creating multi-parameter, multi-scale sensing network without the need for local power source.

# 1.Introduction

The Brillouin optical time domain reflectometry (BOTDR) optical fiber sensor has garnered great attention as a method for distributed monitoring. Since their inception, BOTDR sensors have been extensively researched and applied in various fields, including structural health monitoring, submarine cable surveillance, and pipeline inspection [1-3]. BOTDR operates by injecting light into an optical fiber, where it undergoes spontaneous Brillouin scattering. The resulting frequency shift of the scattered light depends on the temperature and strain experienced by the fiber, enabling the detection of environmental changes through Brillouin Frequency Shift (BFS) measurements [4]. The distributed optical fiber sensor (DOFS) offers the unique advantages of long distance with certain spatial resolution (normally in the meter-scale) and the typical sensing parameters include strain, temperature and vibration. However, for applications like crack detection in concrete structures [5], localized detection at potential risk points is highly desirable. In such cases, high-accuracy point sensors are necessary as complementary methods. Furthermore, in complex environments like deep-sea oil platforms, underground tunnels, or areas prone to geological hazards, besides the DOFS, multiple sensors and multiple parameters of monitoring are required to reduce the false alarm or increase the detection accuracy [6-8]. Although DOFS can operate passively along the sensing region, other sensor types require a continuous power supply to maintain real-time, uninterrupted operation. Deploying a reliable power source in areas with limited infrastructure poses significant challenges, hindering the effective implementation of the sensing network.

Recently, the power-over-fiber (PoF) technology has attracted significant attention for its ability to deliver electrical power over fiber links with high efficiency and stability. Eliminating the need for copper cables and offering immunity to electromagnetic interference, PoF presents a promising power supply solution for remote, hazardous, or difficult-to-reach locations. PoF systems can be deployed across various types of fiber links.

For instance, with large core size and high damage threshold. The Multimode Fiber (MMF) is utilized in distributed antenna systems to transmit information and energy over 300 meters with a power transmission of 250 mW [9]. Subsequently, the transmitted optical power goes up to 9.7 W with a transmission efficiency of ~75% [10]. To enhance the PoF performance, double-cladding fibers (DCF) consisting of the single-mode core, inner cladding and outer cladding have been employed. The electrical power greater than 400 mW is realized in 100 m DCF [11]. Higher optical power of 150 W was successfully delivered over 1 km [12], but the energy-optical transmission efficiency was only 19.7% due to the large coupling loss and transmission loss. Multi-Core Fiber (MCF), a newer type of optical fiber featuring multiple independent cores within a common cladding area, has also been explored. Using a seven-core MCF, fiber-optic information and energy co-transmission was achieved, with three cores dedicated to energy transmission and one core for signal transmission [13]. In 2024, synchronous radio, pump, and power fiber transmission was demonstrated using a single seven-core fiber for downlink radio transmission [14]. In this scheme, each MCF core was limited to a maximum optical input power of 1 W to prevent damage, and a total of 4 W of optical power was transmitted using four cores. To further increase the damage threshold, anti-resonant hollow-core fibers (AR-HCF) are good candidates. The experiments demonstrated near-diffraction-limited beam quality and no fiber damage during 300 W high-power delivery [15]. However, all the above-mentioned fibers are not widely-implemented and cost-effective. Particularly, gas-filling is necessary to realize distributed fiber sensing in AR-HCF, limiting their applicability over long distances [16]. Standard Single-Mode Fiber (SSMF) is widely used in current optical networks and is compatible with various off-the-shelf devices. An SSMF-based Radio over Fiber (RoF) communication system capable of energy co-transmission was proposed to deliver 15.6 mW of electrical power [17]. More recently, a system demonstrated the simultaneous transmission of 10 W optical power and a fiber-carried 5G signal over a 1-kilometer SSMF, highlighting the significant potential of SSMF-based PoF [18]. we present a PoF-BOTDR hybrid system for the first time. In the SSMF link, the power delivery wavelength is set at 1 μm and the sensing wavelength at 1.55 μm, where high-power laser sources and passive devices are readily available. The large wavelength separation minimizes cross-talk between the two channels. Our system successfully delivers up to 6.2 W of optical power while simultaneously performing BOTDR temperature sensing over a 1.3-kilometer fiber link. We found that the high-power laser light is transmitted with an efficiency of up to 65.7%, and it has minimal impact on the Brillouin frequency shift (BFS) of the sensing probe. This work offers a promising solution for deploying multi-sensor networks in remote or hard-to-access areas without the need for local power supplies and demonstrates the feasibility of developing PoF-DOFS hybrid systems.

## 2.Experiments

The experimental setup of the proposed PoF-BOTDR system is illustrated in Figure 1. For the BOTDR part, a narrow linewidth laser (NLL) operating at 1550.12 nm with a linewidth of 1 kHz generates a continuous-wave (CW) optical carrier of 17 dBm. A 50/50 coupler split the signal into two pathways where the upper path signal serves as the optical sensing probe while the lower path functions as the local oscillator (LO). The upper path signal is modulated into optical pulses by a semiconductor amplifier (SOA), with a repetition frequency of 10 kHz and a pulse width of 300 ns. Then it is amplified by a pulsed Erbium-doped fiber amplifier (EDFA) to an optimized power level of 4.4 dBm. An optical bandpass filter excludes the ASE noise generated by the EDFA. In the LO branch, the microwave source generates a swept microwave signal which is carrier-suppressed modulated by the Mach-Zehnder modulator (MZM) for producing frequency-swept optical signal. To enhance the signal-to-noise ratio of the BOTDR system, we use a polarization scrambler (PS) to equalize the polarization interference. The backscattered signal from the circulator is connected to a balanced photodetector (BPD) for optical-electronic conversion and subsequently amplified by an electrical amplifier. Finally, the Brillouin backscattered signal after envelope detection is subjected to data processing. For the PoF part, a single-mode laser diode (SM-HPLD) centered at 1064 nm wavelength is used to generate a high-power light. To combine the sensing and power light, we used a single-mode wavelength-division multiplexer (SM-WDM) specifically designed to handle high power at 1064 nm and 1550 nm. This multiplexer features a high damage threshold near 10 W, ensuring the stability of the system. Although the core diameter of the SM-HPLD pigtail (5.8 µm) is different from that of the SM-WDM (8.2 µm), successful fiber fusion splicing with an insertion loss less than 0.2 dB is achieved. The combined light is then launched into the 1.3 km SMF for transmission. At the output of the SSMF, the residual forward sensing light and high-power power light are separated by using another SM-WDM. The transmitted power light is characterized by using an optical power meter (OPM).

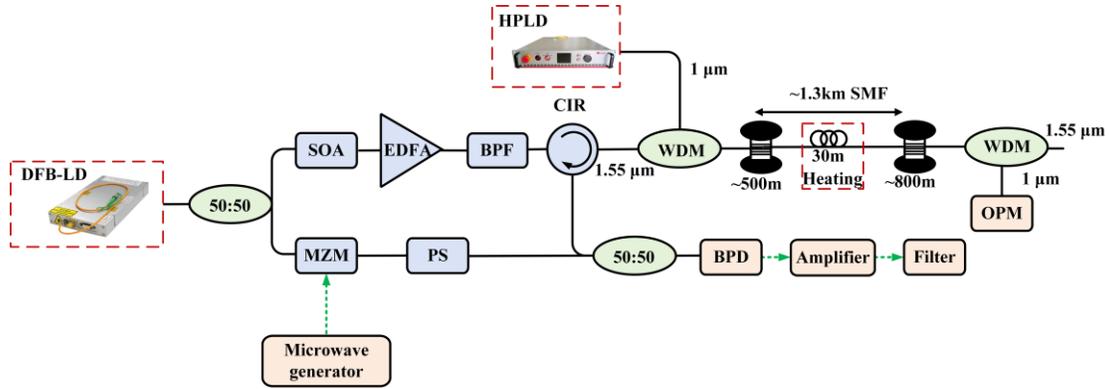

Fig. 1 The experimental setup of the PoF-BOTDR system.

## 3.Results and discussions

Figure 2 shows the collected power of power light and corresponding power delivery efficiency with (w/) and without (w/o) sensing light under different injection powers at room temperature. The power delivery efficiency is defined as the ratio of collected optical power to the injection power of power light. We can conclude that the collected power of power light and corresponding power delivery efficiency with sensing light is almost identical to the condition without sensing light. This indicates that the effect of the BOTDR part on the PoF part is negligible. With 6.24 W power injection, the received power comes to 4.1 W, corresponding to a power delivery efficiency of 65.7%. Considering the InGaAsP-based multi-junction photovoltaic power converter, the optical-to-electrical conversion efficiency is 35% [15] and an electrical power of 1.44 W can be obtained. It can be seen that the collected optical power increases linearly with the injection power, which means both SBS and SRS have been effectively suppressed.

Next, we assess the stability of power light transmission across different temperatures, specifically at 22 °C, 30 °C, 40 °C, 50 °C, and 60 °C. Figure 3 shows the collected power and the corresponding power delivery efficiency of the power light at different injected power, and it can be seen that the proposed PoF-BOTDR system is able to maintain a stable operation in 22°C~60°C environments.

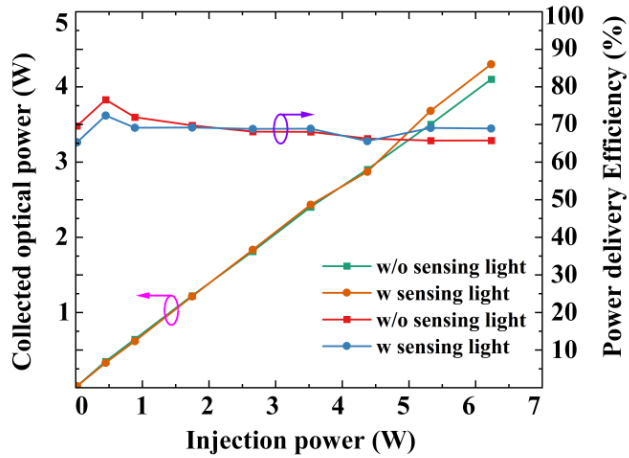

Fig. 2 Collected power of power light and corresponding power delivery efficiency with (w/) and without (w/o) sensing light under different injection powers at room temperature.

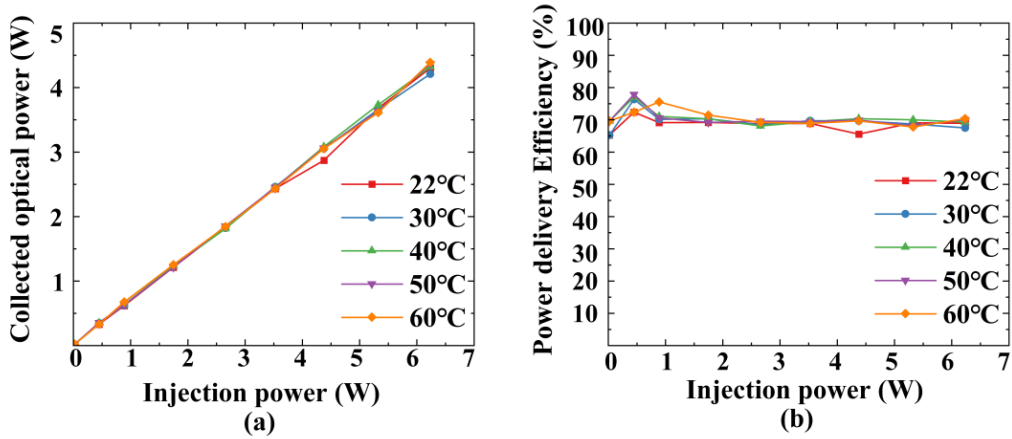

Fig. 3 (a)Collected power of power light and (b)corresponding power delivery efficiency under different injection powers at 22 ℃, 30 ℃, 40 ℃, 50 ℃ and 60 ℃.

Finally, to evaluate the sensing performance of the PoF-BOTDR system in the scenario of optical power transmission, the middle 30 m length of the 1.3 km fiber was submerged a water bath and the rest of the fiber was kept at room temperature. The temperatures in the water bath were set to 30 °C, 40 °C, 50 °C, and 60 °C. Figure 4 shows the BFS along the 1.3 km fiber at different temperatures for a certain injection power. Figure 5 displays BFS for various injection powers at different temperatures. It can be qualitatively seen that the power light has little effect on the temperature sensing performance of the PoF-BOTDR system at the laser output power before WDM range from

1 W to 7 W. For further evidence, we calculated the standard deviation of the BFSs at different temperatures versus room temperature for different input powers, and it can be seen from Figure 6 that the power light has a negligible effect on the temperature measurement performance of the PoF-BOTDR system.

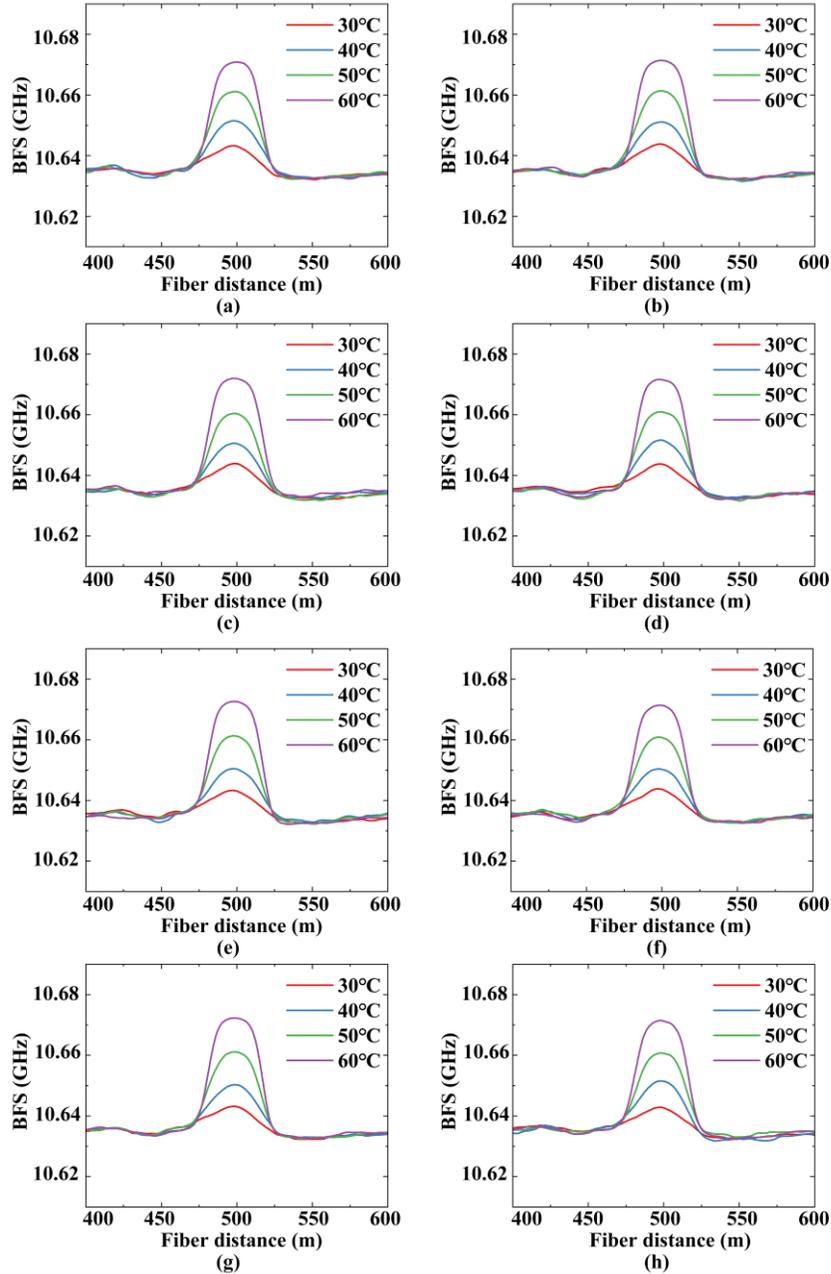

Fig. 4 Brillouin frequency shift at different temperatures for (a)0 W (b)1 W (c)2 W (d)3 W (e)4 W (f)5 W (g)6 W (h)7 W laser output power before WDM.

Taking the water bath temperature of 50°C as an example, when there is no power optical transmission, a temperature change of 28 °C in the central 30-meter section of the fiber results in a Brillouin Frequency Shift (BFS) change of 30.27 MHz, corresponding to a BOTDR temperature sensitivity of 1.08 MHz/°C. When varying the optical power input from 1 W to 7 W in 1 W increments, the BFS changes are 30.70 MHz, 29.37 MHz, 30.19 MHz, 30.58 MHz, 30.06 MHz, 30.75 MHz, and 29.90 MHz, respectively. Consequently, the BFS errors are 0.43 MHz, 0.9 MHz, 0.08 MHz, 0.31 MHz, 0.48 MHz, and 0.37 MHz, with an average temperature error of ±0.39 °C.

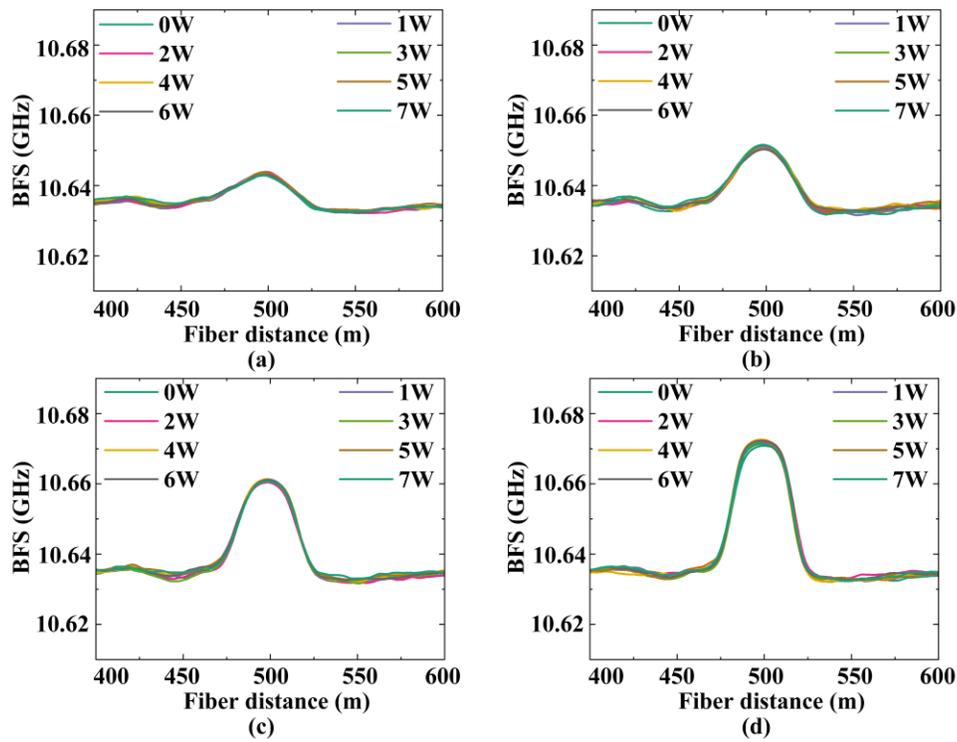

Fig. 5 Brillouin frequency shift with different laser output power before WDM at (a)30 °C(b)40 °C (c)50 °C (d)60 °C.

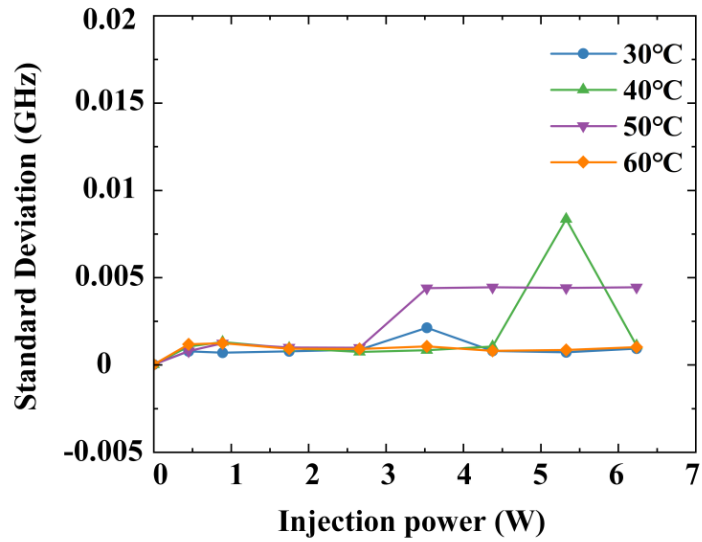

Fig. 6 Standard deviation of BFS at different temperatures versus room temperature for different injection powers

## 4.Conclusion

In conclusion, we demonstrated the PoF-BOTDR hybrid system over a 1.3 km SSMF link for the first time and proved its feasibility by analyzing the power delivery and sensing performance of the system. In this system, the power delivery channel is placed in the 1 μm wavelength, where normal dispersion weakens the modulation instability. The linewidth of the laser is optimized to suppress the SBS as well. The maximum collected optical power is 4.3 W corresponding to a power delivery efficiency of 65.7%. The sensing channel is placed in the 1.55 μm wavelength, which is compatible with the current fiber-optic communication/sensing system. We found that under power delivery background, operation failure will not occur in the BOTDR. The PoF-BOTDR system proposed in this study is expected to offer potential solution for constructing multi-sensor networks in remote or difficult-to-access locations. It also proves the possibility for simultaneously realizing PoF and DOFS with in the same fiber link, which brings wider application and research areas to these two technologies.

**Funding.** The National Natural Science Foundation of China (42327803, 62275097, 62205313); Open Fund of SINOPEC Key Laboratory of Geophysics; Open Project Program of Wuhan National Laboratory for Optoelectronics (2023WNLOKF007);